\documentclass{article}

\usepackage[preprint]{neurips_2026}
\usepackage{subfiles}
\usepackage{amsmath}
\usepackage{graphicx}
\usepackage{makecell}

\usepackage[utf8]{inputenc} % allow utf-8 input
\usepackage[T1]{fontenc}    % use 8-bit T1 fonts
\usepackage{hyperref}       % hyperlinks
\usepackage{url}            % simple URL typesetting
\usepackage{booktabs}       % professional-quality tables
\usepackage{amsfonts}       % blackboard math symbols
\usepackage{nicefrac}       % compact symbols for 1/2, etc.
\usepackage{microtype}      % microtypography
\usepackage{xcolor}         % colors
\usepackage[capitalize]{cleveref}
\usepackage{wrapfig}
\newcommand{\modelname}{\textsc{AttriCite}}
\newcommand{\modelnameours}{\modelname{} (ours)}
\newcommand{\datasetname}{\textsc{CiteAlign}}

\title{\modelname{}: Training an Open 4B Model for Citation Recovery toward Faithful Attribution}
\workshoptitle{AI-Native Academia: Authorship, Peer Review, and Conference Governance under AI}

\author{%
  \textbf{Yee Man Choi\textsuperscript{1,2}},
  \textbf{Xuehang Guo\textsuperscript{3}},
  \textbf{Songcheng Cai\textsuperscript{1}},
  \textbf{Yimu Wang\textsuperscript{1}}, \\
  \textbf{Yi R. (May) Fung\textsuperscript{4}},
  \textbf{Qingyun Wang\textsuperscript{3}},
 \\
 \textsuperscript{1}University of Waterloo,
 \textsuperscript{2}Vector Institute,
 \textsuperscript{3}College of William and Mary,\\
 \textsuperscript{4}University of Illinois Urbana-Champaign\\
 \textsuperscript{1}\{ymchoi,songcheng.cai,yimu.wang\}@uwaterloo.ca \\ \textsuperscript{3}\{xguo15,qwang16\}@wm.edu\\
 \textsuperscript{4}yifung2@illinois.edu
}

\begin{document}

\maketitle

\begin{abstract}
Faithful citation attribution begins with identifying the intended source for a scientific claim. We study this source-identification capability through citation recovery: recovering the paper cited by the original author from a citation-bearing passage. Our evaluation adopts the published author's citation as an observable human attribution signal and uses target recovery as a proxy for progress toward faithful attribution. We introduce \modelname{}, an open 4B-parameter model trained for tool-using citation recovery within the CiteGuard retrieval environment, together with \datasetname{}, a 7,607-instance computer-science dataset drawn from recent scientific literature. For controlled evaluation, we construct a 709-instance benchmark subset of \datasetname{}, comprising 410 development instances from 2024 publications and 299 temporally held-out test instances from 2025 publications. Across three runs at an inference temperature of 0.7, GRPO fine-tuning improves Qwen3-4B from $49.4\%\mathbin{\pm}1.5\%$ to $59.8\%\mathbin{\pm}0.2\%$ target-match accuracy, a gain of 10.4 percentage points. Despite using only 4B parameters, \modelname{} outperforms gpt-oss-20b and comes within 3.9 points of GPT-5.4-mini, while Gemma 4 31B IT achieves the strongest overall performance at $72.0\%\mathbin{\pm}1.0\%$. We release the model and collection pipeline
%(\url{https://anonymous.4open.science/r/AttriCite-C6D5})
(\url{https://github.com/KathCYM/AttriCite})
to support reproducible research on citation recovery toward faithful attribution in a continually evolving scientific literature.
\end{abstract}

\section{Introduction}
Citation attribution is fundamental to the scientific reward system. As Robert K. Merton wrote in his foreword to Eugene Garfield's \emph{Citation Indexing}, ``That claim resides only in the recognition of the source of the contribution by peers'' \citep{garfield1979citation}. Citations ground claims in prior work and credit the researchers who produced it. Preserving this connection is becoming more difficult as the scientific literature grows \citep{larsen2010growth,bornmann2015growth}, a challenge further amplified by the rise of LLM use in scientific writing and its association with greater manuscript production \citep{liang2024mapping,kusumegi2025scientific}.

\begin{figure*}[t]
    \centering
    \includegraphics[width=\textwidth,trim=3mm 3mm 3mm 3mm,clip]{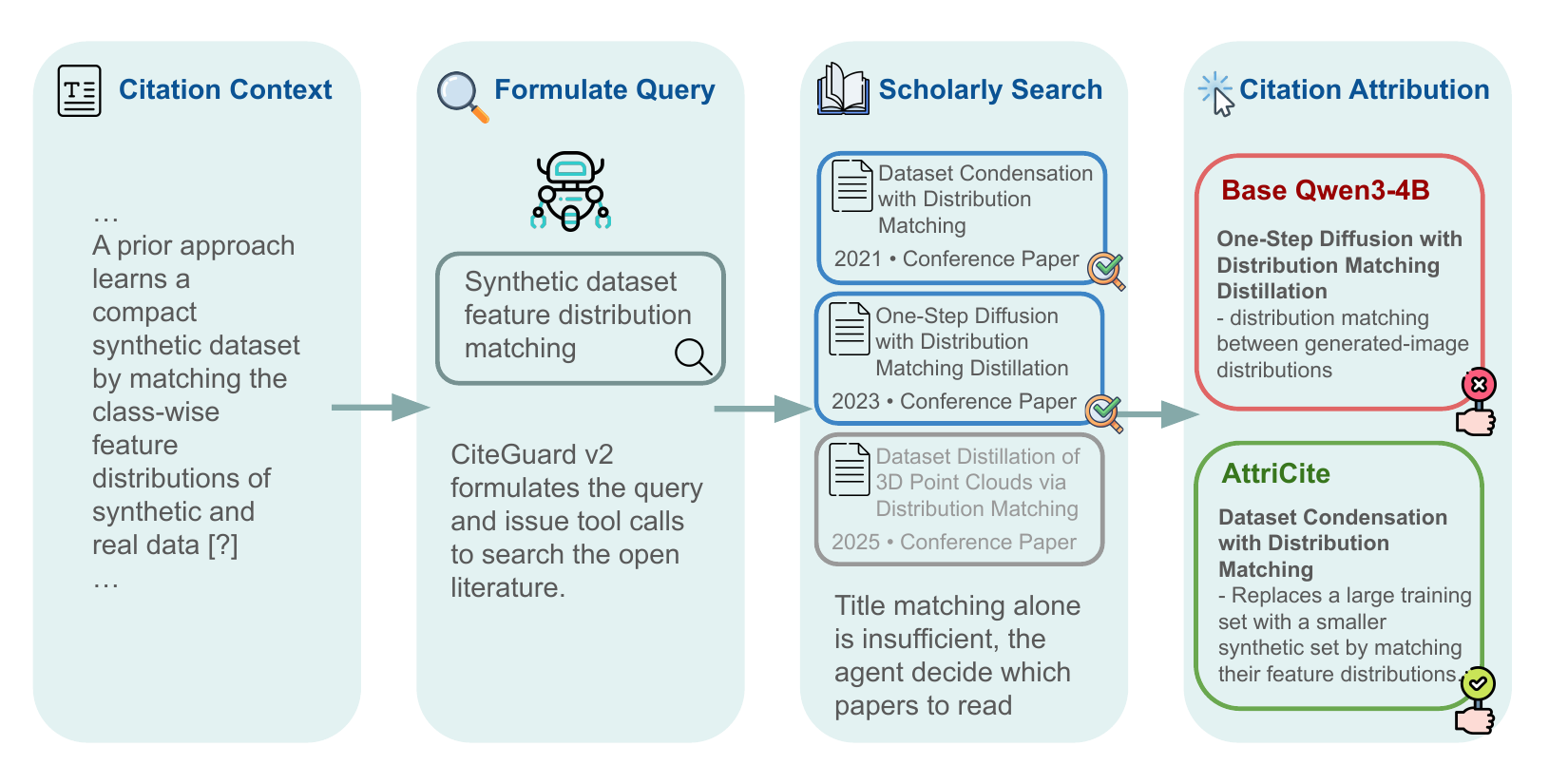}
    \caption{Open-ended citation recovery with \modelname{}. Given a passage with a
    missing citation, the model searches an external scholarly index and identifies
    the source cited by the original author among papers with similar terminology.}
    \label{fig:teaser}
\vspace{-1em}
\end{figure*}

Faithful attribution depends first on correctly identifying the intended scholarly source. This capability is increasingly important because LLMs can generate fabricated or inaccurate references \citep{ji2023hallucination,huang2025hallucination,walters2023fabrication}, and large-scale audits find nonexistent citations in scientific papers \citep{sakai-etal-2026-hallucitation,zhao2026hallucinations}. Source recovery can then be paired with claim--source verification, which assesses evidential support \citep{onweller2026cited}.

This paper targets the recovery component: given a scientific passage with its citation masked, recover the paper cited by the original author. We assume that a citation chosen by a published author provides an observable human signal of intended attribution, and we use agreement with that citation as a proxy for the source-identification component of faithful attribution. \Cref{fig:teaser} illustrates this open-ended process, in which a system formulates queries, inspects retrieved papers, and selects the recorded source. Citation verification evaluates the complementary claim--source pair once a candidate has been discovered \citep{shi2026citeaudit,onweller2026cited}. Open-ended recovery remains challenging even with search tools, with a substantial gap between language models and human researchers \citep{press2024citeme}.
Retrieval-based methods improve attribution \citep{choi2026citeguard}, but repeated search and validation are costly. Existing benchmarks are also static while the literature evolves \citep{press2024citeme,zheng2026citerag}, limiting training and evaluation on newly published work.

We introduce \modelname{}, an open 4B-parameter model trained for citation recovery within the CiteGuard retrieval environment. We fine-tune Qwen3-4B with group relative policy optimization (GRPO) \citep{shao2024deepseekmath} to search scholarly literature and select a source within a bounded interaction budget. Across three runs, fine-tuning improves target-match accuracy from $49.4\%\mathbin{\pm}1.5\%$ to $59.8\%\mathbin{\pm}0.2\%$. Despite using only 4B parameters, \modelname{} outperforms gpt-oss-20b and comes within 3.9 points of GPT-5.4-mini, while Gemma 4 31B IT achieves the strongest overall performance at $72.0\%\mathbin{\pm}1.0\%$.

To support training and evaluation as the scientific literature evolves, we also release \datasetname{}, a citation-recovery dataset constructed through a scalable collection pipeline. Its temporally separated splits evaluate whether systems can recover recent recorded sources by searching an external Semantic Scholar index, and the pipeline enables extension to new publication years and venues.

Our contributions are:
\begin{itemize}
    \item We introduce \modelname{}, an open 4B model trained for tool-using citation recovery as a step toward faithful citation attribution, and show that GRPO fine-tuning yields a 10.4-point target-match gain over the same base model.
    \item We release \datasetname{}, a citation-recovery dataset covering recent scientific literature, with temporally separated splits designed to evaluate retrieval of previously unseen recorded sources.
    \item We release a scalable collection pipeline and separate target retrieval from final selection to identify where fine-tuning improves citation recovery.
\end{itemize}

\section{Problem Formulation}
\label{sec:problem}
\subsection{Citation Recovery toward Faithful Attribution}

Scientific writing uses citations to connect claims to prior work and supporting evidence. We operationalize progress toward faithful citation attribution through \emph{citation recovery}: given a passage and its surrounding context, identify the paper cited by the original author. Our proxy assumption is that the published author's citation provides an observable human signal of the source intended for the passage. The context is important because similar statements may refer to different methods, results, or lines of work. The output is the identity of the recorded source rather than a generated reference string.

Let $x_i$ denote the citation context, $\mathcal{P}$ denote the literature accessible through a scholarly search interface, and $p_i^\ast\in\mathcal{P}$ denote the paper cited in the source document. The evaluated task is to learn a mapping $f_\theta$ such that

\begin{align*}
    f_\theta(x_i) = p_i^\ast.
\end{align*}

This observable label enables automatic evaluation \citep{choi2026citeguard}. Under our proxy assumption, target recovery measures the model's ability to reconstruct a published author's source attribution. This source-identification capability forms the retrieval stage of a broader faithful-attribution pipeline, followed by evidential-support assessment and consideration of alternative valid citations.

We study the \emph{open-ended} setting: the system receives neither a candidate set nor the citing paper's bibliography and must search an external scholarly index. This reflects practical research-agent use, where discovering the source is part of the task. In contrast, citation verification begins with a proposed claim--source pair and asks whether that source supports the claim \citep{shi2026citeaudit,onweller2026cited}.

\subsection{Retrieval Interaction}

We follow the CiteGuard retrieval framework \citep{choi2026citeguard}. At step $k$, the model observes the passage and interaction history $h_k=(x_i,a_1,o_1,\ldots,a_{k-1},o_{k-1})$, where $a_j$ is a search, inspection, context-request, or selection action and $o_j$ is the resulting observation. A policy $\pi_\theta(a_k\mid h_k)$ chooses the next action subject to a budget $B$. The trajectory ends when the model selects a paper or exhausts the budget.

This interaction separates two sources of error. A \emph{retrieval failure} occurs when the target never appears among the retrieved candidates. A \emph{selection failure} occurs when the target is retrieved but the model chooses another paper. The distinction is useful because additional search alone does not guarantee better scientific retrieval or more reliable citations \citep{hu2026sage,onweller2026cited}.

\subsection{Evaluation}

Our primary metric is target-match accuracy for the recorded citation. Let $R_i$ indicate whether target $p_i^\ast$ appears in any candidate set returned during the trajectory. We jointly define end-to-end accuracy, retrieval recall, and conditional selection accuracy as

\begin{align*}
    \mathrm{Accuracy}
    &= \frac{1}{N}\sum_{i=1}^{N}\mathbb{I}[\hat{p}_i=p_i^\ast],
    &\qquad
    \mathrm{Recall}_{\mathrm{ret}}
    &= \frac{1}{N}\sum_{i=1}^{N}R_i, \\
    \mathrm{Accuracy}_{\mathrm{sel}\mid\mathrm{ret}}
    &= \frac{\sum_{i=1}^{N}\mathbb{I}[\hat{p}_i=p_i^\ast]}
            {\sum_{i=1}^{N}R_i}.
\end{align*}

Because the agent can select only a retrieved paper, $\mathrm{Accuracy}=\mathrm{Recall}_{\mathrm{ret}}\times\mathrm{Accuracy}_{\mathrm{sel}\mid\mathrm{ret}}$.

Paper identities are matched first using persistent scholarly identifiers and then, when required by metadata or publication-version variation, using a normalized-title fallback. Retrieval recall measures whether search surfaces the target while conditional selection accuracy measures whether the agent recognizes it once retrieved.

\section{\datasetname{}: Dataset Construction}
\label{sec:citealign} 
Existing citation-recovery benchmarks provide limited training data and often use random splits that can expose evaluation source or target papers during training. We introduce \textbf{\datasetname{}}, a dataset for training and temporally separated evaluation of open-ended citation recovery.

Each instance contains a scientific passage, its source paper, and the paper cited by the original author, but evaluated systems receive only the passage and search Semantic Scholar directly, without a candidate set or oracle retrieval corpus. This setting reflects how research agents operate over an external, continually evolving scholarly index.

\subsection{Collection Pipeline}
\label{sec:collection-pipeline}

We collect papers from the official proceedings of ACL, CVPR, ICLR, ICML, and NeurIPS for 2024 and 2025. The pipeline extracts single-citation passages, replaces the citation marker with \texttt{[CITATION]}, and resolves the corresponding bibliography entry to a canonical Semantic Scholar record.

To ensure that targets are retrievable through the search interface, we retain only instances for which the target appears among the top Semantic Scholar results for at least one query constructed solely from the citation-bearing passage. The successful construction query is used only for filtering and is hidden during evaluation; the agent must formulate its own searches, inspect competing papers, and select the target. We retain passage, bibliography, target-resolution, and search metadata for auditing. Full parsing, resolution, filtering, and audit details appear in \Cref{app:citealign-construction}.

%\subsection{Temporal Splits}
\label{sec:citealign-splits}

We use 2024 instances for training and validation and 2025 instances for testing. Every test target is absent from the complete eligible 2024 pool, not merely from the sampled training set, ensuring evaluation on previously unseen cited works. We additionally remove overlap in normalized source-paper titles, target-paper titles, and exact passages across partitions. We initially sample 350 training, 60 validation, and 300 test instances with equal representation from the five venues. Audit excludes one invalid test passage, leaving 299 test instances. Full sampling and deduplication details appear in \Cref{app:citealign-construction}.

%\subsection{Dataset Composition}

\Cref{tab:citealign-full-composition} summarizes the 7,607-instance collection by venue and year, while \Cref{tab:citealign-statistics} reports the 709-instance subset used in the primary experiments. The primary partitions contain no overlapping source or target papers. Including the separate 143-instance biomedical set, the metadata release contains 7,750 records.

\begin{table}[t]
    \begin{minipage}[t]{0.45\textwidth}
        \caption{CiteAlign composition by source-paper venue and year. The complete collection contains 7,607 instances.}
        \label{tab:citealign-full-composition}
        \centering
        \small
        \begin{tabular}{lrrrr}
            \toprule
            Venue & 2024 & 2025 & Total & Share \\
            \midrule
            ACL     &  225 &  331 &  556 &  7.3\% \\
            CVPR    &  666 & 3,308 & 3,974 & 52.2\% \\
            ICLR    &   82 &  151 &  233 &  3.1\% \\
            ICML    &   87 &   79 &  166 &  2.2\% \\
            NeurIPS & 1,341 & 1,337 & 2,678 & 35.2\% \\
            \midrule
            Total   & 2,401 & 5,206 & 7,607 & 100.0\% \\
            \bottomrule
        \end{tabular}
    \end{minipage}
    \hfill
    \begin{minipage}[t]{0.52\textwidth}
        \caption{CiteAlign primary-experiment subset (350 training, 60 validation, and 299 test instances). The partitions contain no overlapping source or target papers.}
        \label{tab:citealign-statistics}
        \centering
        \small
        \begin{tabular}{lrrr}
            \toprule
            Split & Instances & Source papers & Target papers \\
            \midrule
            Train      & 350 & 319 & 329 \\
            Val &  60 &  60 &  60 \\
            Test       & 299 & 299 & 295 \\
            \midrule
            Total      & 709 & 678 & 684 \\
            \bottomrule
        \end{tabular}
    \end{minipage}
\end{table}

For the exploratory scaling study, we use expanded splits of 1,000 training and 200 validation instances from the same eligible 2024 pool, while retaining the 299-item test set. The expanded partitions remain disjoint in records, source papers, and target papers. Full sampling details appear in \Cref{app:retrieval-shaped-reward}.

\subsection{Biomedical Cross-Domain Set}
\label{sec:biomed-set}

We additionally construct a 143-instance biomedical test set using the same pipeline, drawing from open-access 2025 Europe PMC articles in medical imaging and cancer genomics. After deduplication, no passage, source-paper title, or target-paper title overlaps with the data used to train \modelname{}. We use this set only for out-of-domain evaluation; no biomedical instances are used for training or model selection. Further construction details appear in \Cref{app:biomed-construction}.

\subsection{Quality Control}

Automated checks require each instance to contain one masked citation, an unambiguous bibliography mapping, a resolved target, and a target recoverable through passage-only search. Split-level checks deduplicate identical passages and ensure that source and target papers do not overlap across partitions.

We use an initial 100-example development audit to improve passage extraction, citation--reference mapping, and canonical entity resolution. Before repair, 89 examples require no change, ten require canonical-entity corrections, and one is unusable. After incorporating these findings, we freeze a separate 100-record holdout with no prior adjudication; 99 pass all checks, while one extraction failure is excluded. No reviewed or automatically flagged case remains unresolved in the release. Full audit procedures and accounting appear in \Cref{app:manual-audit,app:holdout-audit,app:full-collection-audit}.

\section{\modelname{}: Reinforcement Learning}
\label{sec:method}
\modelname{} specializes Qwen3-4B for citation recovery in the retrieval environment introduced by CiteGuard \citep{choi2026citeguard}. We retain the existing agent interface and five-action budget rather than introducing a new retrieval framework. Given a citation context and its interaction history, the model formulates searches, inspects retrieved papers, and selects one previously retrieved paper. Requiring the final prediction to correspond to a paper retrieved and resolved through Semantic Scholar grounds every output in an indexed scholarly record, addressing nonexistent-reference hallucinations at the agent level. We treat recovery of the published author's citation as a proxy for the source-identification stage of faithful attribution. The source paper containing the passage is excluded from search results to prevent recovery through its bibliography. \Cref{app:agent-interface} describes the inherited interface.

\subsection{Reinforcement Learning}

We fine-tune all model parameters using group relative policy optimization (GRPO) \citep{shao2024deepseekmath}. For each training passage, we sample $G=8$ agent trajectories, based on pilot experiments and available GPU memory, and assign a binary outcome reward:

\begin{align*}
    r(\tau)=
    \begin{cases}
        1, & \text{if the selected paper matches the recorded target},\\
        0, & \text{otherwise}.
    \end{cases}.
\end{align*}

Matches are determined using Semantic Scholar paper identifiers, with normalized-title matching used only as a fallback for metadata or publication-version variation. Rewards are normalized within each group to produce relative advantages, and the model is optimized with a clipped policy-gradient objective and a KL penalty against the reference policy. Search-service exceptions return an error observation and zero immediate tool reward, remain in the rollout record, and may be followed by another action within the budget.

Because the reward evaluates only the final attribution, it does not prescribe a particular search trajectory. The model can learn when to refine a query, inspect a candidate, or compare papers across searches. Under our compute budget, we conduct a small-data study using 350 \datasetname{} training instances and a disjoint 60-instance validation set for checkpoint and decoding-parameter selection. Full training and interface configurations are reported in \Cref{app:training-config}; an exploratory comparison with target-retrieval reward shaping and a 1,000-instance training set appears in \Cref{app:retrieval-shaped-reward}.

\begin{wrapfigure}{R}{0.48\textwidth}
    \centering
    \includegraphics[width=\linewidth]{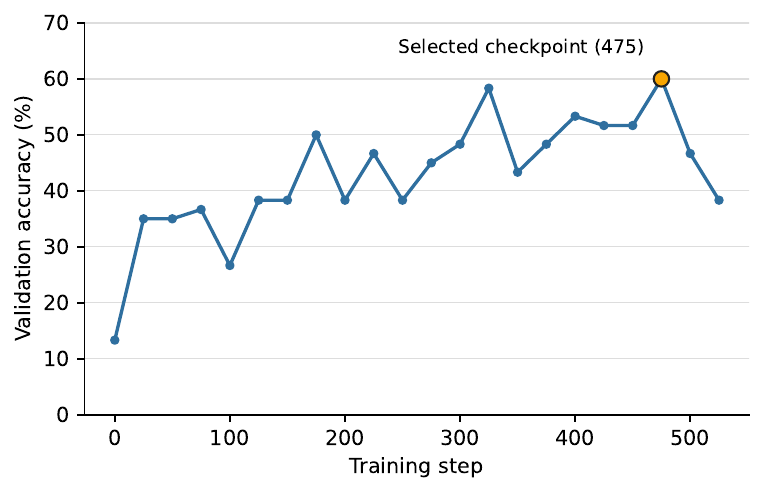}
    \caption{Validation accuracy across training checkpoints. Step 475 is selected for test evaluation.}
    \label{fig:checkpoint-validation}
\vspace{-1em}
\end{wrapfigure}

\paragraph{Training dynamics.}
We evaluate 22 checkpoints at 25-step intervals on the 60-instance validation set. As shown in \Cref{fig:checkpoint-validation}, validation accuracy is noisy because of the small validation set and stochastic agent trajectories. Step 475 attains the highest validation accuracy and is selected before any test evaluation; all evaluator and decoding choices are likewise frozen before scoring the test set. Selecting the maximum among 22 noisy evaluations can yield validation optimism, so we use validation only for checkpoint selection and base our performance claims on the untouched test set.

\subsection{Evaluation Setup}

We evaluate on the 299-instance \datasetname{} test set. Its passages come from 2025 publications, and every target paper is absent from the complete eligible 2024 pool used to construct the training and validation data. We report mean accuracy across three runs.

This temporal separation prevents overlap with our task-specific training data, but it does not guarantee that a pretrained model has never encountered a test paper, an earlier preprint, or its metadata. Comparisons between base Qwen3-4B and \modelname{} remain controlled because they share the same pretrained checkpoint; comparisons across model families may additionally reflect differences in pretraining data and prior exposure to the evaluated papers.

Our controlled comparisons use the same CiteGuard environment, prompts, retrieval backend, and action budget. We compare base Qwen3-4B, GRPO-trained Qwen3-4B (\modelname{}), the open gpt-oss-20b and Gemma 4 31B IT models, and proprietary models including GPT-5.4-mini and Claude Haiku 4.5. The controlled base-model comparison estimates the total effect of our GRPO training procedure relative to the unchanged checkpoint; the present study does not decompose that effect against matched supervised objectives. Comparisons using a different retrieval backend are identified as non-controlled.

To examine cross-domain generalizability from the computer-science venues represented in \datasetname{} to biomedicine, we compare base Qwen3-4B and \modelname{} on the 143-instance biomedical set described in \Cref{sec:biomed-set}. No biomedical examples are used for training or model selection.

\paragraph{Metrics.}
Our primary metric is \emph{target-match accuracy}: the proportion of instances for which the selected paper resolves to the recorded target. The finalized evaluator matches Semantic Scholar paper identifiers first and uses normalized title similarity only as a fallback for metadata and publication-version variation; manually adjudicated ambiguous title pairs override the threshold. A deterministic, threshold-focused review of 100 title pairs validates this fallback policy (\Cref{app:evaluation-details}). Under our proxy assumption, the metric measures recovery of the source attributed by the published author and serves as an automatic indicator of progress on the source-identification component of faithful attribution. We also report inference failures and the number of tool actions.

To locate the source of performance differences, we report the retrieval and conditional-selection metrics defined in \Cref{sec:problem}. Implementation details and qualifications for this diagnostic appear in \Cref{app:evaluation-details}.

\paragraph{Protocol.}
We run each model three times at inference temperature 0.7. The exception is Claude Haiku 4.5, which we evaluate once on \datasetname{} because of API cost and omit from the biomedical evaluation. For models with three completed runs, we report the mean, standard deviation, and two-sided 95\% confidence interval using the appropriate $t$ distribution; for Claude, we report single-run accuracy. Biomedical retrieval uses Semantic Scholar's Medicine field filter. Predictions in both domains are scored by the same identifier-first entity matcher with the frozen title fallback. For the controlled base--\modelname{} comparison, we additionally average correctness per test item across the three runs and apply a paired Bayesian bootstrap over the 299 shared items. This yields a posterior over the accuracy difference that accounts for finite-test-set uncertainty while preserving the paired design. Search dates and returned paper identifiers are recorded because Semantic Scholar is an evolving external index.

\section{Experimental Results and Analysis}
\label{sec:experiments}
\label{sec:results} 
\subsection{Main Results}

\begin{figure*}[t]
    \centering
    \includegraphics[width=0.9\textwidth]{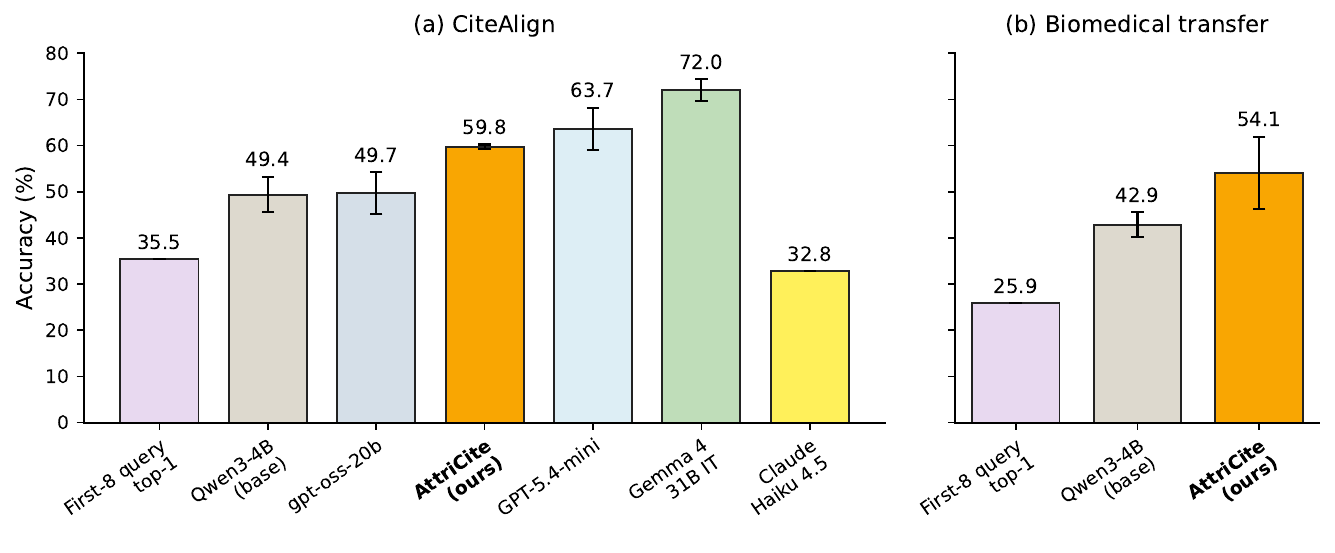}
    \caption{Target-match accuracy on (a) \datasetname{} and (b) the biomedical cross-domain set. In both panels, the deterministic construction baseline submits the top-ranked Semantic Scholar result for a fixed query comprising the first eight non-stopword passage tokens and has no error bar. Biomedical runs use the Medicine field filter. All results use identifier-first entity matching with the frozen title fallback. Model error bars show two-sided 95\% $t$-intervals over three runs; Claude Haiku 4.5 is evaluated once.}
    \label{fig:main-results}
\end{figure*}

\Cref{fig:main-results} compares performance on \datasetname{} and the biomedical cross-domain set, while \Cref{tab:main-results} reports target-match results for the main evaluation.

\begin{table}
    \caption{Target-match accuracy on \datasetname{} at temperature 0.7. The deterministic construction baseline uses the top-ranked result for a fixed first-eight-content-token query. For models, we report three-run means, standard deviations, and two-sided 95\% $t$-intervals, except for Claude Haiku 4.5, which is evaluated once because of API cost.}
    \label{tab:main-results}
    \centering
    \small
    \begin{tabular*}{\textwidth}{@{\extracolsep{\fill}}lccc@{}}
        \toprule
        System & Accuracy & SD & 95\% CI \\
        \midrule
        First-8 query top-1 & 35.45 & -- & -- \\
        Qwen3-4B (base) & 49.39 & 1.54 & 45.55--53.22 \\
        gpt-oss-20b & 49.72 & 1.84 & 45.15--54.30 \\
        Gemma 4 31B IT & \textbf{72.02} & \textbf{0.97} & \textbf{69.62--74.42} \\
        GPT-5.4-mini & 63.66 & 1.84 & 59.08--68.23 \\
        Claude Haiku 4.5 (1 run) & 32.78 & -- & -- \\
        \textbf{\modelnameours{}} & 59.75 & 0.19 & 59.28--60.23 \\
        \bottomrule
    \end{tabular*}
\end{table}

Across three runs, the base Qwen3-4B agent achieves $49.39\%\mathbin{\pm}1.54\%$ accuracy, whereas GRPO-trained \modelname{} achieves $59.75\%\mathbin{\pm}0.19\%$. Fine-tuning therefore yields an absolute improvement of 10.37 percentage points. A paired Bayesian bootstrap over the 299 test items gives a posterior mean gain of 10.37 points and a 95\% credible interval of 6.36--14.49 points. The posterior probability that \modelname{} improves over the base model is greater than 0.999, and the probability that the gain exceeds five points is 0.996. All three \modelname{} runs outperform all three base-model runs, showing that the gain is not driven by a single replicate. GPT-5.4-mini reaches $63.66\%\mathbin{\pm}1.84\%$, outperforming \modelname{} by 3.90 points. The larger open gpt-oss-20b model obtains $49.72\%\mathbin{\pm}1.84\%$. Gemma 4 31B IT achieves the highest accuracy at $72.02\%\mathbin{\pm}0.97\%$, exceeding \modelname{} by 12.26 points. Claude Haiku 4.5 obtains 32.78\% in a single run covering all 299 test instances; 145 trajectories terminate with a valid paper selection and 154 terminate with an error outcome. We report this result for completeness and exclude it from substantive model comparisons; \Cref{app:evaluation-details} gives the outcome accounting. All results use the final adjudicated 299-item test set and frozen entity-matching policy. Together, these results show that task-specific training allows a 4B model to outperform some larger open models and come within 3.90 points of GPT-5.4-mini.

\subsection{Cross-Domain Generalization}

\begin{table}
    \caption{Target-match performance on the 143-instance biomedical cross-domain set using Semantic Scholar's Medicine field filter and the same identifier-first entity matcher with frozen title fallback used for the main set. The deterministic First-8 baseline is a single frozen run; model results are summarized across three runs.}
    \label{tab:biomedical-results}
    \centering
    \small
    \begin{tabular*}{\textwidth}{@{\extracolsep{\fill}}lccc@{}}
        \toprule
        Model & Accuracy & SD & 95\% CI \\
        \midrule
        First-8 query top-1 & 25.87 & -- & -- \\
        Qwen3-4B (base) & 42.89 & 1.07 & 40.24--45.54 \\
        \textbf{\modelnameours{}} & 54.08 & 3.15 & 46.25--61.91 \\
        \bottomrule
    \end{tabular*}
\end{table}

Because the policy is trained only on \datasetname{}'s computer-science papers, the biomedical set tests transfer beyond the training domain. With the domain-appropriate Medicine filter and the shared identifier-first evaluator, the fixed First-8 baseline obtains 25.87\%. \modelname{} improves accuracy from $42.89\%\mathbin{\pm}1.07\%$ for base Qwen3-4B to $54.08\%\mathbin{\pm}3.15\%$, a gain of 11.19 points, and exceeds the fixed-query baseline by 28.21 points. This result provides preliminary evidence that the learned policy transfers to the evaluated medical-imaging and cancer-genomics collections, while the lower absolute accuracy quantifies the remaining domain gap.

\subsection{Retrieval and Final Selection}

\begin{table}
    \caption{Retrieval and selection performance pooled across three runs at temperature 0.7. The deterministic baseline uses its single top-ranked result as the selector, so retrieval at rank one and correct selection coincide.}
    \label{tab:retrieval-selection}
    \centering
    \small
    \begin{tabular*}{\textwidth}{@{\extracolsep{\fill}}lccc@{}}
        \toprule
        Model & Retrieval recall & Selection $\mid$ retrieved & Accuracy \\
        \midrule
        First-8 query top-1 & 35.45 & -- & 35.45 \\
        Qwen3-4B (base) & 58.64 & 84.22 & 49.39 \\
        \textbf{\modelnameours{}} & 68.12 & 87.73 & 59.75 \\
        GPT-5.4-mini & 70.90 & 89.78 & 63.66 \\
        \bottomrule
    \end{tabular*}
\end{table}

As a non-agentic construction baseline, we issue one fixed query comprising the first eight non-stopword tokens after removing the citation marker and submit the first non-source-paper result returned by \texttt{search\_relevance}. This top-1 selector recovers 106 of 299 main-set targets, yielding 35.45\% retrieval recall@1 and end-to-end target-match accuracy. With the Medicine filter and the same identifier-first evaluator, it recovers 37 of 143 biomedical targets, or 25.87\%. Conditional selection is not separately applicable because the single retrieved item is the submitted prediction.

As a secondary diagnostic, we reconstruct target retrieval from both the structured candidate lists and the raw observations following snippet-search actions, including trajectories that end in error. Relative to base Qwen3-4B, \modelname{} raises retrieval recall by 9.48 percentage points, from 58.64\% to 68.12\%, and selection accuracy conditional on retrieval by 3.51 points, from 84.22\% to 87.73\%. Their products recover the corresponding pooled end-to-end accuracies of 49.39\% and 59.75\%. Snippet histories add five target retrievals for the base model and none for \modelname{} across the three runs. The larger change in retrieval recall is consistent with fine-tuning primarily improving the agent's ability to surface the recorded target, accompanied by a smaller improvement in selecting it once found.

Relative to GPT-5.4-mini, the remaining gaps are 2.78 points in retrieval recall and 2.05 points in conditional selection. The directly computed end-to-end difference is 3.90 points. Reconstruction details and qualitative examples of retrieval gains, selection gains, and residual errors appear in \Cref{app:evaluation-details,app:qualitative-examples}.

\section{Related Work}
\paragraph{Citation Recommendation and Attribution}
Citation recommendation is commonly formulated as retrieval and ranking over a fixed scholarly corpus. Prior work uses manuscript or citation-context representations \citep{bhagavatula2018content,medic2020improved}, citation-informed paper embeddings \citep{cohan2020specter,ostendorff2022scincl}, and evidence-grounded, reconstruction, or retrieval-generation methods for local citation prediction \citep{ghoshroy2024ilciter,celik2025citebart,zheng2026citerag}. These methods assume a predefined corpus or index. CiteME and CiteGuard instead evaluate recovery of an author-recorded citation from an open search space, with CiteGuard adding iterative search and candidate validation \citep{press2024citeme,choi2026citeguard}. We build on this interaction setting by training a compact open policy and introducing recent, temporally separated training and evaluation data.

\paragraph{Citation Verification}
Citation verification asks whether a reference supports its associated claim. SciFact introduced expert-annotated claims with supporting or refuting evidence, and SciFact-Open extended verification to retrieval over 500,000 abstracts \citep{wadden2020scifact,wadden2022scifactopen}. More recent evaluations study citations in generated answers, including citation correctness and completeness in ALCE \citep{gao2023alce}, multi-stage claim and evidence verification in CiteAudit \citep{shi2026citeaudit}, and link validity, relevance, and factual support in deep-research reports \citep{onweller2026cited}. Verification is complementary to our setting: \modelname{} recovers the author-recorded source, whereas verification determines whether that source supports the claim.

\paragraph{Scientific Literature Agents}
Scientific agents retrieve literature and synthesize citation-backed answers \citep{lala2023paperqa,asai2024openscholar}, while STORM uses iterative retrieval and perspective-guided questioning for grounded long-form generation \citep{shao2024storm}. Recent benchmarks evaluate reasoning-intensive scientific retrieval and constrained target-paper discovery \citep{hu2026sage,xiong2026autoresearchbench}. These systems assess broader research workflows in which quality depends on retrieval, coverage, organization, and support. In contrast, \datasetname{} isolates recovery of a single author-recorded citation, enabling automatic target-match evaluation and separate analysis of retrieval and final selection.

\paragraph{Learning to Use Search Tools}
Language models can learn to interleave reasoning, tool use, and multi-turn search through prompting, self-supervision, and outcome-based reinforcement learning \citep{yao2023react,schick2023toolformer,jin2025searchr1,song2025r1searcher}. We use GRPO \citep{shao2024deepseekmath} with a target-recovery reward, training the policy to search, inspect results, and select a specific scholarly record under a bounded action budget. Unlike open-domain QA rewards, our terminal reward directly measures recovery of the target scholarly entity. Our controlled base-model comparison estimates the overall effect of this training procedure, without separately comparing against supervised trajectory learning.

\label{sec:related}
\section{Limitations}
\label{sec:limitations}
\paragraph{Dataset Scope and Construction.}
\datasetname{} covers five computer-science venues and two publication years, while the biomedical set provides an initial out-of-domain evaluation in two areas. Automatic construction may introduce parsing, resolution, or citation-span errors. The discoverability filter further defines the benchmark population around targets reachable through Semantic Scholar.

\paragraph{Evaluation Proxy.}
We assume that a citation chosen by a published author represents the intended source attribution and use target recovery as a proxy for this component of faithful attribution. This proxy focuses on agreement with the recorded citation; broader evaluation can incorporate evidential-support judgments and alternative appropriate sources.

\paragraph{Resource and responsible use.}
\modelname{} is intended to assist source discovery. Incorrect attribution can deny credit or lend a claim unwarranted authority, so users should inspect retrieved evidence before publication. Additional qualifications concerning pretraining exposure, the live retrieval backend, the reward, and computational resources appear in \Cref{app:evaluation-details}.

\paragraph{Learning objective.}
The controlled base-model comparison estimates the effect of our complete GRPO training procedure, not the necessity or superiority of GRPO. We do not include a matched supervised fine-tuning or behavioral-cloning baseline built from successful tool trajectories. Such a comparison is needed to distinguish the benefits of outcome-based reinforcement learning from those of task-specific trajectory training more generally.

\section{Conclusion}
We introduced \modelname{}, an open 4B citation-recovery model, and \datasetname{}, a recent dataset with a scalable collection pipeline. GRPO fine-tuning improves Qwen3-4B by 10.4 target-match points, with a paired Bayesian 95\% credible interval of 6.4--14.5 points. \modelname{} outperforms gpt-oss-20b and comes within 3.9 points of GPT-5.4-mini; retrieval analysis associates most of the gain with surfacing the target paper. The biomedical evaluation provides preliminary evidence of transfer to medical imaging and cancer genomics. \modelname{} establishes a reproducible improvement in open-ended source identification; claim--source support verification remains the complementary next stage.

%\section*{References}

{
\small
\bibliographystyle{plain}
\bibliography{reference}
%[1] Alexander, J.A.\ \& Mozer, M.C.\ (1995) Template-based algorithms for
%connectionist rule extraction. In G.\ Tesauro, D.S.\ Touretzky and T.K.\ Leen
%(eds.), {\it Advances in Neural Information Processing Systems 7},
%pp.\ 609--616. Cambridge, MA: MIT Press.

%[2] Bower, J.M.\ \& Beeman, D.\ (1995) {\it The Book of GENESIS: Exploring
%  Realistic Neural Models with the GEneral NEural SImulation System.}  New York:
%TELOS/Springer--Verlag.

%[3] Hasselmo, M.E., Schnell, E.\ \& Barkai, E.\ (1995) Dynamics of learning and
%recall at excitatory recurrent synapses and cholinergic modulation in rat
%hippocampal region CA3. {\it Journal of Neuroscience} {\bf 15}(7):5249-5262.
}

%%%%%%%%%%%%%%%%%%%%%%%%%%%%%%%%%%%%%%%%%%%%%%%%%%%%%%%%%%%%

\appendix
\section{Additional Dataset and Experimental Details}
\label{app:additional-details}

\subsection{\datasetname{} Construction Details}
\label{app:citealign-construction}

\paragraph{Proceedings collection.}
We collect paper titles, landing pages, PDF URLs, venues, and publication years from official ACL Anthology, CVF, PMLR, NeurIPS, and ICLR proceedings pages. Venue-specific crawlers produce a common source-paper manifest. PDFs are cached, failures are recorded, and collection state is checkpointed after each paper.

\paragraph{PDF and citation parsing.}
The pipeline extracts PDF text, separates the main body from the bibliography using detected \emph{References} or \emph{Bibliography} headings, and segments the body into sentences. It supports parenthetical and narrative author--year citations as well as numeric citations. We retain sentences linked to one reference, replace their citation marker with \texttt{[CITATION]}, and discard cases that cannot be mapped unambiguously or do not contain exactly one marker.

\paragraph{Reference resolution.}
For each bibliography entry, we query Semantic Scholar using an inferred title, an author-based query, and a compact form of the full reference. We retrieve up to ten results per query and score each candidate as

\begin{align*}
    s_{\textit{resolve}}(r,p)
    = 0.7\,s_{\textit{token}}(r,p)
    + 0.3\,s_{\textit{sequence}}(r,p)
    + 0.05\,\mathbb{I}[y_p \in r],
\end{align*}

where $s_{\textit{token}}$ measures title-token overlap, $s_{\textit{sequence}}$ is normalized string similarity, and the final term rewards year agreement. We retain the highest-scoring result when $s_{\textit{resolve}} \geq 0.55$. Searches are restricted to computer-science papers published no later than the source paper.

\paragraph{Discoverability filtering.}
We independently search using only the citation-bearing passage. Query variants use the first 14 tokens, the first eight non-stopword tokens, and a local window of up to six content tokens on either side of the citation. An instance is retained when its resolved target occurs within the top 20 Semantic Scholar relevance results for at least one query. Titles match when their normalized string similarity is at least $0.90$. This query and its target rank are retained for auditing but are hidden from the evaluated model.

\paragraph{Split construction.}
After normalizing titles and passages, we remove exact duplicate triples of passage, source title, and target title. Test instances are sampled from 2025 only when their targets are absent from every eligible 2024 instance. We then remove from the 2024 pool any instance whose passage, source title, or target title overlaps with test. We sample 82 instances per venue, holding out 12 for validation and using 70 for training. Sampling greedily prioritizes source and target papers used least often, with a seeded stable hash for tie-breaking; validation similarly favors papers absent from training. We find no exact passage overlap with the benchmark used in the original CiteGuard release.

\paragraph{Audit records.}
For each retained instance, we store the raw passage, original citation marker, bibliography entry, resolved Semantic Scholar identifier, canonical title and year, resolution query and score, and successful discoverability query and rank. Candidate rows and audit records are written incrementally, allowing interrupted runs to resume and new venues or publication years to use the same downstream construction stages.

\subsection{Full-Collection Audit and Release Accounting}
\label{app:full-collection-audit}

The release audit processes all 7,754 construction records across 13 collection files. Deterministic checks flag possible passage-extraction failures, citation--bibliography mismatches, source-paper self-resolutions, canonical-entity errors, malformed contexts, duplicates, and questionable claim--source relationships. Flagged records are resolved using the source PDF, its bibliography, and cross-index paper records; corrections are stored as explicit adjudications rather than silently modifying the raw collection output.

All cases flagged by the strengthened audit are adjudicated before release: recoverable records receive traceable corrections and unrecoverable contexts are rejected. After the separate held-out audit excludes one additional extraction failure, the final metadata contains 7,750 records: 7,607 computer-science and 143 biomedical. The released validation report records the detailed decision accounting, confirms that no flagged case remains unresolved, and verifies zero duplicate release identifiers and zero missing passage fingerprints or experimental-split source-PDF URLs. This accounting describes adjudication coverage for automatically flagged cases rather than claiming manual inspection of every unflagged record.

\subsection{Development Manual Verification and Adjudication}
\label{app:manual-audit}

We manually verify a 100-instance sample drawn from the original 300-instance test set. To balance venue coverage, we select 20 examples from each of ACL, CVPR, ICLR, ICML, and NeurIPS. Selection is deterministic: examples within each venue are ordered by a SHA-256 hash of the dataset identifier and the fixed seed \texttt{citealign-audit-20260828}. The audit therefore does not depend on model predictions or automatic resolution confidence. This stratified design checks every venue equally; because the complete collection is CVPR-heavy, its unweighted 1/100 exclusion rate is an audit-sample result rather than an estimate of collection-wide error prevalence.

For each sampled example, the reviewer examines the unmasked source passage in its source paper, the citation marker, the matched bibliography entry, and the resolved target record. The annotation form separately records whether (i) the passage was extracted from the paper body, (ii) the citation marker maps to the intended bibliography entry, (iii) the bibliography entry resolves to the correct canonical paper, and (iv) the cited source bears a direct, indirect, background, or unclear relationship to the surrounding claim. It also records whether alternative sources could reasonably fit the passage, overall label validity, reviewer confidence, an error category, and free-text adjudication notes.

The adjudicated relationship labels comprise 68 direct-support cases, five partial-support cases, 20 background attributions, six related-only cases, and one unclear case. Reviewers identify a plausible alternative source for 15 passages and no plausible alternative for 85. These categories characterize how the recorded citation relates to its passage; they do not alter the target-recovery label.

We use these annotations to improve construction, not only to estimate residual error. In particular, the review motivates checks that reject bibliography fragments as passages, prevent a citation from resolving to its own source paper, validate citation markers against the matched bibliography entry, and preserve verified canonical replacements as explicit adjudications. We add regression tests for these failure modes and apply the revised audit-and-adjudication procedure to the complete construction pool. Consequently, the released collection code incorporates the review findings, and the released metadata contains the corrected targets rather than the original erroneous resolutions.

Review findings are used as development feedback for the released pipeline rather than as a held-out estimate of an earlier implementation. Reviewers verify canonical entities against the source bibliography and replace identifiers, titles, URLs, and years when the bibliographic evidence establishes a correction. The review also identifies one NeurIPS example whose extracted ``passage'' is itself a bibliography entry and therefore provides no valid citation-recovery context. We exclude this record instead of assigning a replacement target, reducing the released test set from 300 to 299 examples.

Before adjudication, 89 audited examples require no change, ten have an incorrect canonical target entity, and one is an extraction failure whose context is a bibliography entry. After pipeline improvement, repair, and adjudication, all 99 audited records retained in the release pass quality control, while the unusable context is excluded, yielding an audit-sample exclusion rate of 1/100. The audit trail preserves record identifiers, reasons, and correction authority so that each adjudication remains traceable. The subsequent full-collection audit described in \Cref{app:full-collection-audit} applies the strengthened procedure beyond this development sample.

\subsection{Finalized-Pipeline Holdout Audit}
\label{app:holdout-audit}

After freezing the improved pipeline and full-collection adjudications, we draw a second deterministic sample of 100 records using the SHA-256 seed \texttt{attricite-finalized-holdout-20260830}. The sampling frame contains 6,868 non-experimental records and excludes every training, validation, main-test, and biomedical-test record, all records from the development audit, and every record previously sent to full-dataset adjudication. This separation prevents the holdout from providing feedback to the reported experiments or reusing cases that motivated the pipeline changes. The sample follows the collection distribution rather than balancing venues: it contains 61 CVPR, 32 NeurIPS, five ACL, and two ICLR records.

The reviewer applies the same four checks for passage extraction, citation--reference mapping, canonical entity resolution, and claim--source relationship. Ninety-nine records pass all checks. One record has the correct citation mapping and canonical target, but its extracted sentence is interrupted by figure text before the citation context is complete; we mark it as an extraction failure and exclude it from the public metadata after freezing the audit decision. The held-out failure rate is therefore 1/100, and all 99 retained holdout records pass review. The deterministic sample, completed annotation fields, failure rationale, and summary are included with the released audit artifacts.

\subsection{Biomedical Cross-Domain Set Construction}
\label{app:biomed-construction}

We retrieve open-access 2025 articles with available PDFs through Europe PMC using two topical collections: medical imaging and cancer genomics. The \datasetname{} pipeline extracts single-reference passages, resolves their targets through Semantic Scholar, and applies the same passage-only discoverability requirement used for the computer-science collection. This yields 51 medical-imaging and 93 cancer-genomics candidates. We remove one exact duplicate, producing 143 evaluation instances from 59 unique source papers and 141 unique target papers.

We compare normalized passages, source titles, and target titles against the available CiteGuard and \datasetname{} training files, including the 350-instance training split and the larger eligible training pools. No overlap is found. The released validation report records input counts, duplicate removal, overlap checks, journal composition, and the maximum repetitions per source and target paper.

\subsection{Agent Interface and Prompts}
\label{app:agent-interface}

\paragraph{Prompt construction.}
Training and local evaluation use the same prompt builder. The system message defines citation recovery, supplies the JSON response contract and tool-use constraints, recommends short query reformulations when a search fails, and includes one worked tool-use example. The user message consists of the following instruction followed by the citation-bearing passage:

\begin{quote}
\small
``You are now given an excerpt. Find me the paper cited in the excerpt, using the tools described above. Please make sure that the paper you select really corresponds to the excerpt: there will be details mentioned in the excerpt that should appear in the paper. If you read an abstract and it seems like it could be the paper we are looking for, read the paper to make sure. Also: sometimes you will read a paper that cites the paper we are looking for. In such cases, please go to the references to find the full name of the paper we are looking for, search for it, and then select it.''
\end{quote}

The model sees the masked passage but not the source-paper title, source year, or target identity. The environment uses the source title and year only to exclude the citing paper and restrict searches to papers available no later than the source publication. Each model turn must contain exactly one JSON object with a short \texttt{reason} and an \texttt{action}. The action contains a \texttt{name} and only the arguments required by that action: \texttt{query}, \texttt{paper\_id}, or \texttt{paper\_title}.

\paragraph{Actions and observations.}
The inherited CiteGuard interface exposes seven actions:

\begin{description}
    \item[\texttt{search\_relevance}] returns up to ten Semantic Scholar results ranked by relevance;
    \item[\texttt{search\_citation\_count}] retrieves up to 100 relevance matches, reranks them by citation count, and returns the top ten;
    \item[\texttt{search\_text\_snippet}] returns up to ten matching snippets with paper titles and section labels;
    \item[\texttt{read}] downloads and extracts the open-access PDF for a paper in the latest search buffer;
    \item[\texttt{find\_in\_text}] returns sentences containing a requested string from a paper in the latest search buffer;
    \item[\texttt{ask\_for\_more\_context}] records a focus query; in our fixed benchmark it returns a deterministic message instructing the model to continue from the original passage; and
    \item[\texttt{select}] chooses any paper returned by an earlier paper search and terminates the trajectory.
\end{description}

Structured search observations contain the paper identifier, title, an abstract truncated to 1,000 characters, and citation count. Full observations are capped at 16,000 characters. The latest-search restriction for \texttt{read} and \texttt{find\_in\_text} prevents arbitrary identifier access, while \texttt{select} may refer to any paper accumulated during the trajectory. The environment terminates after a valid \texttt{select} or after five actions. Invalid actions receive an error observation and still consume the action budget. 

\paragraph{Released specification.}
The anonymous release contains the complete, unabridged prompt and worked example in \texttt{src/retriever/prompt\_templates/few\_shot\_tool.txt}; the shared user instruction in \texttt{src/retriever/prompt\_config.py}; the injected JSON contract and dataset conversion in \texttt{training/verl/prepare\_dataset.py}; and the executable schema, buffer rules, observation formatting, and reward behavior in \texttt{training/verl/citeguard\_tool.yaml} and \texttt{training/verl/citeguard\_tool.py}. This makes the appendix description auditable without reproducing the long demonstration verbatim.

\subsection{Training Configuration}
\label{app:training-config}

We train all parameters using veRL with FSDP and asynchronous vLLM rollouts. The actor uses gradient checkpointing, and the base model's generic thinking mode is disabled. Training and evaluation use the same structured action protocol and bounded interaction history.

\begin{table}
    \caption{\modelname{} training configuration. We optimize all Qwen3-4B parameters with GRPO under the same five-action budget used at evaluation.}
    \label{tab:training-config}
    \centering
    \small
    \begin{tabular*}{\textwidth}{@{\extracolsep{\fill}}lr@{}}
        \toprule
        Configuration & Value \\
        \midrule
        Base model & Qwen3-4B \\
        Optimization & Full-parameter GRPO \\
        Training epochs & 3 \\
        Rollouts per prompt & 8 \\
        Prompt batch size & 2 \\
        Learning rate & $1\times10^{-6}$ \\
        KL coefficient & $1\times10^{-3}$ \\
        Rollout temperature / top-$p$ & 0.7 / 0.95 \\
        Maximum agent actions & 5 \\
        Maximum context length & 32,768 tokens \\
        Training hardware & 4$\times$ NVIDIA L40S \\
        Time to selected checkpoint & 6 h 27 min wall-clock \\
        Active step time to selected checkpoint & 6.05 h \\
        Peak allocated memory & 40.4 GiB per GPU \\
        Peak reserved memory & 42.9 GiB per GPU \\
        \bottomrule
    \end{tabular*}
\end{table}

\subsection{Exploratory Target-Retrieval Shaping Comparison}
\label{app:retrieval-shaped-reward}

We conduct a supplementary comparison between the terminal outcome reward used for the primary model and a target-retrieval-shaped variant. The outcome reward credits only a correct final selection. The shaped variant retains that terminal reward and adds a positive bonus when the target paper appears in any search result during the trajectory,
\begin{align*}
    r_{\mathrm{shaped}} = r_{\mathrm{select}} + \lambda\,\mathbb{I}[\text{target retrieved}],
\end{align*}
where the retrieval event is recorded by the tool environment and $\lambda=0.1$, supplied through the released \texttt{SEARCH\_HIT\_REWARD} configuration. This is reward shaping based on an observable retrieval event, not supervision of latent reasoning steps. We compare the rewards at two data scales: the 350-instance training subset with 60 validation instances and the 1,000-instance training set with 200 validation instances. Within each scale, the outcome and shaped runs use the same non-reward configuration. The subset schedule uses three epochs, two prompts and eight rollouts per optimizer update, validation and checkpointing every 25 updates, and 525 scheduled updates; the selected outcome and shaped checkpoints occur at steps 475 and 425, after nominally 7,600 and 6,800 sampled trajectories. The expanded schedule likewise uses three epochs, two prompts and eight rollouts per update, with validation and checkpointing every 50 updates and 1,500 scheduled updates; its selected outcome and shaped checkpoints occur at steps 350 and 450, after nominally 5,600 and 7,200 trajectories. The expanded runs use a 24,576-token per-GPU optimization limit and parameter and optimizer offloading, whereas the subset runs use 32,768 tokens without actor offloading. Consequently, matched reward comparisons within a scale isolate the shaping choice, but comparisons across scales also vary data, selected exposure, checkpoint frequency, memory settings, and compute. This analysis was conducted after the primary configuration was established and is exploratory rather than a replacement for the main result.

\begin{table}[t]
    \caption{Exploratory outcome- versus target-retrieval-shaped reward comparison on the 299-instance main test set. Accuracy is the mean across three temperature-0.7 runs with sample standard deviation. Retrieval recall and selection conditional on retrieval are pooled across the same runs. All quantities use the identifier-first evaluator with the frozen title fallback.}
    \label{tab:retrieval-shaped-reward}
    \centering
    \small
    \begin{tabular*}{\textwidth}{@{\extracolsep{\fill}}llccc@{}}
        \toprule
        Training set & Reward & Accuracy & Retrieval recall & Selection $\mid$ retrieved \\
        \midrule
        None & Base model & $49.39\mathbin{\pm}1.54$ & 58.64 & 84.22 \\
        Subset & Retrieval-shaped & $51.95\mathbin{\pm}0.19$ & 62.54 & 83.07 \\
        Subset & Outcome & $\mathbf{59.75\mathbin{\pm}0.19}$ & \textbf{68.12} & \textbf{87.73} \\
        1,000 & Retrieval-shaped & $58.31\mathbin{\pm}2.18$ & 67.22 & \textbf{86.73} \\
        1,000 & Outcome & $\mathbf{58.97\mathbin{\pm}0.84}$ & \textbf{68.67} & 85.88 \\
        \bottomrule
    \end{tabular*}
\end{table}

We align the 299 test items across three runs per system and average each item's binary outcomes before paired inference. A 100,000-sample Bayesian bootstrap over items gives the subset outcome-minus-shaped accuracy difference as 7.80 points (95\% interval 3.49--12.16); a two-sided paired randomization test gives $p=0.00074$. At the 1,000-instance scale, the corresponding difference is 0.67 points (95\% interval $-2.14$--3.57; $p=0.702$), providing no evidence that either reward is superior. Moving from the subset configuration to the expanded retrieval-shaped configuration improves accuracy by 6.35 points (95\% interval 2.33--10.43; $p=0.00264$), whereas the corresponding outcome-reward change is $-0.78$ points (95\% interval $-3.98$--2.54; $p=0.685$). Because the cross-scale configurations differ beyond dataset size, these contrasts describe configuration changes rather than isolated data-scaling effects. The subset outcome advantage appears in both retrieval and conditional selection (nominal paired $p=0.0248$ and $p=0.0259$). We apply Holm correction jointly to four accuracy hypotheses---the two within-scale reward contrasts and the two within-reward cross-scale contrasts---and the subset retrieval and conditional-selection hypotheses. The subset outcome-over-shaped accuracy contrast and the cross-scale retrieval-shaped accuracy contrast remain significant after correction; neither component-level contrast remains significant, and the 1,000-instance reward contrast and outcome-reward cross-scale contrast provide no evidence of a difference.

At the tested coefficient, target-retrieval shaping performs worse than the terminal outcome reward on the subset and does not outperform it at the expanded scale. Its higher score under the expanded configuration shows that the poor subset result is not stable across the two tested configurations, but the experiment does not establish a benefit from shaping. Because the cross-scale settings and selected checkpoint exposures differ and only one shaping coefficient is examined, the comparison should not be interpreted as a general conclusion for or against reward shaping.

\subsection{Models, Services, and Licenses}
\label{app:assets}

\Cref{tab:model-assets} records the exact model identifiers used in our experiments and their applicable licenses or service terms. Qwen3-4B, gpt-oss-20b, and Gemma 4 31B IT are distributed under Apache 2.0; \modelname{} is derived from Qwen3-4B and is released under Apache 2.0. GPT-5.4-mini and Claude Haiku 4.5 were accessed through their providers' APIs under the OpenAI Services Agreement and Anthropic Commercial Terms, respectively. The scholarly-search backend was used for non-commercial research under the Semantic Scholar API License Agreement.

\begin{table}
    \caption{Model and service identifiers used in the experiments. Hosted models are governed by the provider terms listed here rather than an open-weight license.}
    \label{tab:model-assets}
    \centering
    \small
    \begin{tabular*}{\textwidth}{@{\extracolsep{\fill}}lll@{}}
        \toprule
        Reported name & Exact identifier & License or terms \\
        \midrule
        Qwen3-4B & \texttt{Qwen/Qwen3-4B} & Apache 2.0 \\
        \modelnameours{} & step-475 derivative of Qwen3-4B & Apache 2.0 \\
        gpt-oss-20b & \texttt{openai/gpt-oss-20b} & Apache 2.0 \\
        Gemma 4 31B IT & \texttt{google/gemma-4-31B-it} & Apache 2.0 \\
        GPT-5.4-mini & \texttt{gpt-5.4-mini-2026-03-17} & OpenAI API terms \\
        Claude Haiku 4.5 & \texttt{claude-haiku-4-5-20251001} & Anthropic API terms \\
        Scholarly search & Semantic Scholar API & AI2 API agreement \\
        Released code & CiteGuard/\modelname{} implementation & MIT \\
        Dataset layer & \datasetname{} metadata and annotations & CC BY 4.0 \\
        Reconstruction tool & Passage-reconstruction code & MIT \\
        \bottomrule
    \end{tabular*}
\end{table}

The applicable terms are available from the Qwen3-4B model card (\url{https://huggingface.co/Qwen/Qwen3-4B}), gpt-oss-20b model card (\url{https://huggingface.co/openai/gpt-oss-20b}), Gemma 4 31B IT model card (\url{https://huggingface.co/google/gemma-4-31B-it}), OpenAI Services Agreement (\url{https://openai.com/policies/services-agreement/}), Anthropic Commercial Terms (\url{https://www.anthropic.com/legal/commercial-terms}), Together AI Terms of Service (\url{https://www.together.ai/terms-of-service}), and Semantic Scholar API License Agreement (\url{https://api.semanticscholar.org/license/}). The CC BY 4.0 license applies only to metadata and annotations created by us and does not relicense third-party paper text. The public release therefore contains source-paper identifiers, passage locations, and integrity fingerprints rather than source excerpts. An MIT-licensed script reconstructs the passages locally from the original publications.

\subsection{Evaluation Details}
\label{app:evaluation-details}

We measure selection rate, inference errors, and the mean numbers of actions, searches, and paper-inspection calls. Invalid or unparsed actions are recorded separately. For live-search experiments, we also retain the query date and returned Semantic Scholar paper identifiers.

For paired item-level uncertainty, we align the 299 retained test identifiers across the three reported base and trained runs and average each model's binary correctness per item. We apply a paired Bayesian bootstrap by drawing 200,000 Dirichlet$(1,\ldots,1)$ weight vectors over the shared test items using seed 20260828 and computing the weighted mean of the paired item differences. We report the posterior mean, equal-tail 95\% credible interval, probability of any improvement, and probability of an improvement exceeding five percentage points.

Total wall-clock duration to the selected checkpoint was 6 hours 27 minutes on four NVIDIA L40S GPUs, or 25.8 L40S GPU-hours; the training steps themselves account for the 6.05 active hours reported in \Cref{tab:training-config}. Local \datasetname{} evaluation of base Qwen3-4B and \modelname{} used four L40S GPUs with four-way tensor parallelism; biomedical evaluation used two L40S GPUs with two-way tensor parallelism. We report three 299-instance main-set evaluations after audit exclusion and three 143-instance biomedical evaluations for each local model. The gpt-oss-20b and Gemma baselines were served by Together AI, GPT-5.4-mini by OpenAI, and Claude Haiku 4.5 by Anthropic; provider-managed hardware and memory are not exposed to API users and are not local requirements for reproduction.

Across the three \datasetname{} runs at temperature 0.7, mean end-to-end latency is 4.87 seconds per instance for base Qwen3-4B and 2.75 seconds for \modelname{}; the corresponding medians are 3.22 and 2.27 seconds. These measurements include model generation, agent execution, and live Semantic Scholar requests, and therefore depend on server load, network conditions, and the external search service. They should not be interpreted as isolated model-inference latency. The tensor-parallel evaluation setup does not establish the minimum memory required for single-GPU deployment.

Retrieval recall is reconstructed from the candidate lists saved after relevance and citation-count searches together with paper titles parsed from the raw observations following snippet-search actions. This reconstruction covers errored trajectories and avoids the legacy \texttt{is\_in\_search} field, which omits retrievals preceding an error and represents a top-ranked match with the integer zero. Candidate identifiers are matched first; the same adjudicated normalized-title fallback used for final selection handles snippet results without identifiers and publication-version variation. We retain the retrieval--selection decomposition as a secondary diagnostic because it describes where observed errors occur rather than identifying a unique causal mechanism.

\subsection{Qualitative Retrieval and Selection Examples}
\label{app:qualitative-examples}

We derive the following examples from the saved trajectories using deterministic behavioral criteria across all three reported base and \modelname{} runs. The first requires that the base model never retrieve the target while \modelname{} selects it correctly in every run; the second requires both systems to retrieve the target but only \modelname{} to select it correctly in every run; and the third requires \modelname{} to retrieve but never select the target. Titles and actions below illustrate the first run, while the stated outcome patterns hold across all three runs.

\paragraph{Retrieval gain.}
For test instance 2000063, the passage states that the decision to refuse a harmful request is ``mediated by a single direction,'' and the target is \emph{Refusal in Language Models Is Mediated by a Single Direction}. The base model begins with the broad passage query but then drifts to \texttt{decision making cognitive psychology neuroscience}; it never retrieves the target and selects an unrelated resource-rational decision-making paper. \modelname{} instead searches \texttt{decision refuse harmful request single direction} and immediately selects the target. This example illustrates how retaining the claim's distinctive relation and terminology can determine whether the source enters the candidate set.

\paragraph{Selection gain after successful retrieval.}
For instance 2000150, the passage names AdaFuse as a multiview-aware joint detector. Both systems search for \texttt{AdaFuse} and retrieve the target, \emph{AdaFuse: Adaptive Multiview Fusion for Accurate Human Pose Estimation in the Wild}. The base model nevertheless selects the similarly named \emph{AdaFuse: Adaptive Temporal Fusion Network for Efficient Action Recognition}, whereas \modelname{} selects the multiview pose-estimation paper. The retrieval is therefore shared; the improvement comes from aligning the passage's ``multiview'' and ``joint detector'' cues with the correct candidate.

\paragraph{Residual version ambiguity.}
Instance 2000072 attributes improved GPU-memory efficiency to FlashAttention, with \emph{FlashAttention-2: Faster Attention with Better Parallelism and Work Partitioning} recorded as the target. Although the target is present in the retrieved candidates, both base Qwen3-4B and \modelname{} select the earlier \emph{FlashAttention: Fast and Memory-Efficient Exact Attention with IO-Awareness} in all three runs. This is a genuine selection error under entity-level evaluation and shows why matching a method family is insufficient when the passage cites a particular version.

The candidate-selection script and originating result identifiers are retained with the analysis artifacts, allowing these examples to be regenerated from the complete trajectories.

The title fallback is evaluated on a deterministic 100-pair stress sample drawn from the predictions of the systems and runs in the primary main-table comparison and concentrated around the 0.80 decision boundary. The sample includes every fuzzy-only pair that would otherwise receive credit in that primary comparison. This exhaustive primary-output adjudication is used to define frozen overrides; it is not an independent held-out validation sample. The records retain originating result filenames, so we conservatively treat adjudication as non-blinded to system provenance; aggregate model scores were recomputed only after the decisions were frozen. Manual review finds 92\% agreement with the threshold alone and identifies seven false-positive near-title pairs and one legitimate title-change false negative. These decisions are frozen as explicit overrides, while exact Semantic Scholar identifiers remain authoritative. Every reported computer-science prediction and retrieval trace is then scored with the same identifier-first policy and title fallback, including the later exploratory reward variants, without changing the matching rule or overrides. The audit validates the fallback decisions observed in the primary comparison; the frozen rule, rather than a new system-specific audit, is applied to the exploratory variants.

\begin{table}[t]
    \caption{Entity-matching pathways aggregated over the primary main-table test runs.
    ``Exact title'' applies only after identifier mismatch; positive and negative
    overrides are manually adjudicated ambiguous pairs. Every fuzzy-only pair
    credited in these primary runs was reviewed.}
    \label{tab:matcher-pathways}
    \centering
    \small
    \setlength{\tabcolsep}{4pt}
    \begin{tabular}{lrrrrr}
        \toprule
        System
        & Identifier
        & \makecell{Exact\\title}
        & \makecell{Positive\\override}
        & \makecell{Negative\\override}
        & \makecell{Other\\incorrect} \\
        \midrule
        Qwen3-4B (base), 3 runs & 430 & 5  & 8  & 21 & 433 \\
        \modelname{}, 3 runs    & 515 & 12 & 9  & 16 & 345 \\
        gpt-oss-20b, 3 runs     & 432 & 4  & 10 & 23 & 428 \\
        Gemma 4 31B IT, 3 runs  & 618 & 11 & 17 & 7  & 244 \\
        GPT-5.4-mini, 3 runs    & 547 & 13 & 11 & 26 & 300 \\
        Claude Haiku 4.5, 1 run & 94  & 1  & 3  & 3  & 198 \\
        \bottomrule
    \end{tabular}
\end{table}

The construction baseline uses one fixed query function from the released collector: the first eight non-stopword passage tokens after removing the citation marker. We issue it through \texttt{search\_relevance}, use the Computer Science field filter for \datasetname{} and the Medicine filter for the biomedical set, exclude the citing paper, and submit the first remaining result. We freeze the complete returned candidate lists and selected identifiers for every example. Under the shared identifier-first entity policy with frozen title fallback, the selected paper matches the adjudicated target for 106/299 main-set examples (35.45\%) and 37/143 biomedical examples (25.87\%). The released script and response files make both snapshots reproducible and auditable.

Claude Haiku 4.5 is evaluated on all 299 test instances in its single run. Of these trajectories, 145 terminate with a valid paper selection and 154 terminate with an error outcome: 79 contain structured-output format errors, 68 exhaust the action budget, and 7 contain no complete JSON object. Accuracy conditional on a valid selection is 67.59\%; because this conditions on the trajectory outcome, it is not directly comparable to the 32.78\% end-to-end accuracy over all test instances.

%%%%%%%%%%%%%%%%%%%%%%%%%%%%%%%%%%%%%%%%%%%%%%%%%%%%%%%%%%%%

%\newpage
%\input{checklist.tex}

\end{document}